\documentclass[sigconf]{acmart}

\usepackage{stfloats}

\usepackage{microtype}
\usepackage{balance}
\usepackage{booktabs}
\usepackage{graphicx}
\usepackage{multirow}
\usepackage[figurename=Figure]{caption}

\usepackage{caption}
\AtBeginDocument{%
  \providecommand\BibTeX{{%
    \normalfont B\kern-0.5em{\scshape i\kern-0.25em b}\kern-0.8em\TeX}}}

\copyrightyear{2024}
\acmYear{2024}
\setcopyright{rightsretained}
\acmConference[CHI '24]{Proceedings of the CHI Conference on Human Factors in Computing Systems}{May 11--16, 2024}{Honolulu, HI, USA}
\acmBooktitle{Proceedings of the CHI Conference on Human Factors in Computing Systems (CHI '24), May 11--16, 2024, Honolulu, HI, USA}
\acmDOI{10.1145/3613904.3642806}
\acmISBN{979-8-4007-0330-0/24/05}

\acmSubmissionID{7631}

\title[Enabling Parent-Robot Collaboration to Support Children's Learning Experiences]{``It's Not a Replacement:'' Enabling Parent-Robot Collaboration to Support In-Home Learning Experiences of Young Children}

\author{Hui-Ru Ho}
\email{hho24@wisc.edu}
\orcid{0009-0000-3701-2521}
\affiliation{%
  \institution{Dept. of Educational Psychology\\University of Wisconsin--Madison}
  \streetaddress{Department of Educational Psychology, University of Wisconsin--Madison}
  \city{Madison}
  \state{WI}
  \country{USA}
}

\author{Edward Hubbard}
\email{emhubbard@wisc.edu}
\orcid{0000-0002-2701-415X}
\affiliation{%
  \institution{Dept. of Educational Psychology\\University of Wisconsin--Madison}
  \streetaddress{Department of Educational Psychology, University of Wisconsin--Madison}
  \city{Madison}
  \state{WI}
  \country{USA}
}

\author{Bilge Mutlu}
\email{bilge@cs.wisc.edu}
\orcid{0000-0002-9456-1495}
\affiliation{%
  \institution{Dept. of Computer Sciences\\University of Wisconsin--Madison}
  \streetaddress{Department of Computer Sciences, University of Wisconsin--Madison}
  \city{Madison}
  \state{WI}
  \country{USA}
}

\begin{document}

\begin{abstract} 
Learning companion robots for young children are increasingly adopted in informal learning environments. Although parents play a pivotal role in their children's learning, very little is known about how parents prefer to incorporate robots into their children's learning activities. We developed prototype capabilities for a learning companion robot to deliver educational prompts and responses to parent-child pairs during reading sessions and conducted in-home user studies involving 10 families with children aged 3--5. Our data indicates that parents want to work with robots as collaborators to augment parental activities to foster children's learning, introducing the notion of \textit{parent-robot collaboration}. Our findings offer an empirical understanding of the needs and challenges of parent-child interaction in informal learning scenarios and design opportunities for integrating a companion robot into these interactions. We offer insights into how robots might be designed to facilitate parent-robot collaboration, including parenting policies, collaboration patterns, and interaction paradigms. 
\end{abstract}



\begin{teaserfigure}
  \includegraphics[width=\textwidth]{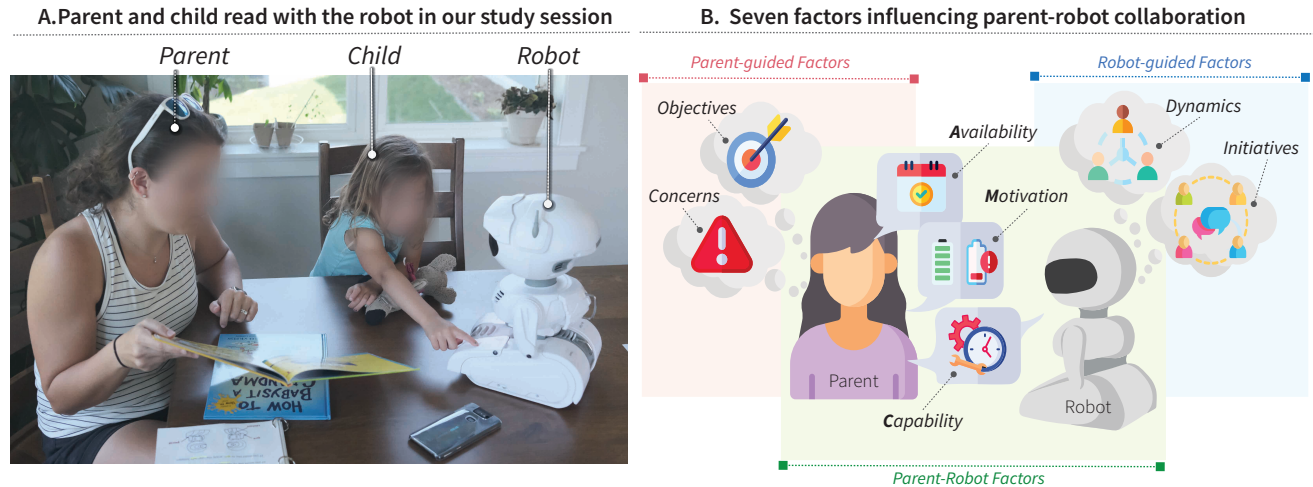}
  \caption{We explored how parents used and envisioned a learning companion robot to support young children's learning at home and introduce the notion of \textit{parent-robot collaboration} as a mechanism for successful integration of robots into the home.}
  \label{fig:individual}
\end{teaserfigure}

\maketitle




\section{Introduction}

As care robots have become more common and entered into formal and informal learning environments, a substantial literature has discussed how social robots could be designed to support children in school-based learning \cite{lopez2018robotic, davison2020working, kennedy2016social, gordon2016affective}, home-based learning \cite{michaelis2018reading, michaelis2019supporting, ho2021robomath, ho2023designing}, as well as various other aspects of life such as well-being \cite{henkel2020robotic} and daily routines \cite{chan2017wakey}. Most of the existing research has primarily concentrated on interaction design between a robot and a single child \cite{kennedy2016social, tanaka2009use, chen2020teaching, ali2019can} or a robot with multiple children \cite{escobar2022robot}, with little work starting to consider the potential aspect of including a parent in the interaction \cite{chan2017wakey, chen2022designing}. 

Recent field deployments of learning companion robots suggest that the use of these robots in a home environment usually involves complex social dynamics. \citet{michaelis2023off} conducted a four-week in-home deployment study and found that a social robot naturally elicited involvement of multiple family members who all engaged with the robot in their individual ways. They recommended offering parents regular reports about children's learning activities with the robot. Moreover, established work in developmental psychology has shown that parents or primary caregivers play critical roles in promoting social and intellectual development, especially in young children \cite{blevins2016early, duncan2007school, peck1992parent}. Not only the frequency, but also the quality, of parental involvement in children's learning can significantly influence children's life outcomes and overall well being \cite{harris2008parents, sui1996effects, elliott2017understanding}. The quality of parental participation in children's learning activities also differ from family to family based on individual characteristics of parents \cite{dowker2021home, elliott2017understanding}.

We would therefore like to know what kind of social dynamics might emerge in interactions between a learning companion robot and parent-child pairs, as well as how parents perceive and envision their use of a learning companion robot to support their children's learning activities at home. Without a thorough understanding of how parents make use of a learning companion robot at home, there is a dearth of design choices for integrating parents into existing child-robot interaction paradigms. It is therefore crucial to understand how social robots could be designed to work together with individual parents or caregivers, and thereby support young children's learning by empowering parents in their strengths, supplementing their weaknesses with assistance from the robot, and minimizing parental concerns via ethical design processes.

In this work, which is part of a larger project on designing learning companion robots, we conducted field studies and interviews with families in their homes using a learning companion robot as a technology probe designed to facilitate reading and math learning. We discussed with parents how they perceive their lived experience interacting with a social robot and their child, and how they envision such technology being designed to support their implementation of parenting strategies within an educational context. 

We found a significant interest in managing parental involvement in children's learning activities to the extent that parents saw the robot as a \textit{collaborator} with them. In addition, we present themes related to parents' preferences and vision of their collaboration with the robot, including factors related to their own parenting policies, factors related to robots' interaction management, and determining factors for the distribution of work between the parent and the robot. We later explain how these themes interconnect and the resulting design considerations for future robots that support parent-child pairs for in-home learning experiences. 

\section{Related Work}

\subsection{Learning Companion Robots} \label{sec-2.1}

Social robots represent a growing technology that could provide social interactions and physically embodied presence. These affordances have the potential to convert an individual learning experience into a socially-enriched one \cite{michaelis2018reading}. Compared to virtual agents, physically embodied robots can similarly offer adaptive assistance \cite{ramachandran2019personalized, leyzberg2014personalizing, schodde2017adaptive} and verbal communications \cite{brown2014positive}, yet distinctly support learners' engagement with the physical world and elicit higher-level social behaviors during learning \cite{belpaeme2018social}, thereby promote better learning outcomes \cite{leyzberg2012physical}.  Prior work has shown that social robots can effectively support school-based learning in mathematics \cite{lopez2018robotic}, science \cite{davison2020working}, and literacy \cite{kennedy2016social, gordon2016affective}. Similarly, they can enhance children's learning during home-based activities such as reading \cite{michaelis2018reading, michaelis2019supporting}, number board games \cite{ho2021robomath}, and math-oriented conversations with parents \cite{ho2023designing}.

Prior work has demonstrated how social robots could support children's learning within varying social dynamics and stakeholders. We can categorize these studies based on the number and type of agents that are involved into interactions involving (1) a robot and a single child, (2) a robot and multiple children, (3) a robot, multiple children, and an adult, (4) a robot, a child, and an adult. First, a substantial body of work has focused on how a social robot could support \textit{a single child}. For example, \citet{kennedy2016social} demonstrated how children learn elements of a second language from the robot in a short-term dyadic interaction context; on the other hand, \citet{tanaka2009use} showed how a child can take the role of a teacher to instruct the robot as a novice, gaining self-confidence and improving learning outcomes. \citet{chen2020teaching} suggested an adaptive role-switching model to flexibly accommodate a child's personalized needs for being a tutor or a tutee dynamically in a one-on-one interaction with a social robot during vocabulary learning. In addition, beyond the context of structured learning,  \citet{ali2019can} demonstrated how an autonomous social robot can foster a child's creativity using verbal and non-verbal behaviors. 

Secondly, some studies have addressed how a social robot interacts with \textit{multiple children} in educational contexts. For instance, \citet{escobar2022robot} investigated how children interact with a social robot in pairs during a problem solving task and found individual differences in the perceived role of the robot based on varying degrees of robot expressivity, suggesting the importance of considering individual perception when designing the robot to support children's learning in groups. In addition, \citet{kanda2012children} uncovered how social behavior exhibited by a robot supported the initial building of relationships among children for collaborative learning in a classroom setting. 

Adding an adult to each of the dynamics mentioned above results in the third and the fourth forms of social dynamics with a social robot. The third form of interaction with a robot involves \textit{multiple children and an adult} who could be a teacher using the robot in a classroom setting \cite{you2006robot}. For instance, \citet{you2006robot} introduced a humanoid robot into a English class as a teaching assistant to support the teacher motivating a group of children. The fourth form of interaction with a robot involves \textit{a child and an adult} who could be a parent \cite{gvirsman2020patricc, chan2017wakey, cagiltay2020investigating, chen2022designing}. Although several studies describe parental perspectives on incorporating a social robot in their child's learning, most of them have not integrated a situated interaction experience with a social robot into their methods, relying instead on surveys or interviews facilitated by pictures of social robots. These studies suggested that parents hold a positive attitude towards the potential of integrating a social robot in their children's learning either at home or school \cite{tolksdorf2020parents, lee2008elementary, oros2014children, louie2021desire}. However, they also reported several concerns including technical challenges \cite{tolksdorf2020parents}, undesirable replacement of the teacher's role \cite{lee2008elementary}, and disruption of children's school-based learning \cite{louie2021desire}. A few studies we found incorporated a physical robot in study sessions with a parent and a child to induce insight into this interaction dynamic \cite{gvirsman2020patricc, chan2017wakey, chen2022designing}. \citet{gvirsman2020patricc} developed a robotic system called Patricc that promoted parent-toddler-robot (toddlers aged 1--3 years old) triadic interaction more than a tablet in a 8-minute learning session, measured by video-coded gaze. Another study designed a technology-based intervention, called WAKEY, for facilitating morning routines of preschoolers (aged 3--6 years old). In a week-long evaluation study, WAKEY was found to mitigate parental frustration about communication regarding morning routines with children and make positive impact on parents' attitude in parenting and communication \cite{chan2017wakey}. In addition, \citet{chen2022designing} designed a platform that integrated a Jibo robot and an Android tablet to facilitate parent-child co-reading activities in a 3--6 week long deployment study, targeting families with children aged 3--7 years old. They uncovered that social robots could improve various aspects of parent-child interaction, recommending that the robot's role in a group setting should be explored from individual group members' perspective and that these opinions should be further integrated in the design of the robot.

\subsection{Parent Involvement in Children's Learning} \label{sec-2.2}

Parents are one of the most crucial socialization agents that influence the acquisition and development of a broad spectrum of cognitive and social skills in children \cite{blevins2016early, duncan2007school, peck1992parent}. Extensive literature in the field of educational psychology has consistently demonstrated the positive impact of parental engagement in children's learning on their academic achievements \cite{fehrmann1987home, hill2004parent}. Moreover, active parental involvement within the home environment holds particular promise for enhancing learning outcomes, surpassing the effects observed in a school setting \cite{harris2008parents, fantuzzo2004multiple, sui1996effects}. Furthermore, \citet{hoffner2002parents} identified a correlation between the extent of parental guidance and children's age, with younger children eliciting more parental mediation compared to their older counterparts. These findings underscore the importance of promoting parental involvement, especially within the home, in the education of young children.

\subsection{Parent-mediated Technology Use} \label{sec-2.3}

As various forms of technology and media become increasingly prevalent, parents are increasingly taking an active role in managing their children's use of these technologies. Parental mediation is critical for young children, as it further determines how children's learning and development is induced by the media they are interacting with. \citet{warren2001words} defined  \textit{parental mediation} as ``strategies parents use to control, supervise or interpret media content for children.'' Parents mediate children's use of technology using different strategies depending on various factors such as their confidence in using technology \cite{dias2016role}. 

\textit{Parental mediation theory} has been continuously expanded by the Human-Computer Interaction (HCI) community to explain the involvement of parents in children's technology and media usage in a broader context. The foundational version of parental mediation theory originated from the patterns of watching television, encompassing three primary strategies: restrictive mediation, active mediation, and co-use \cite{valkenburg1999developing}. Restrictive mediation refers to how parents make rules to guide or prohibit the viewing of certain content; active mediation (sometimes called ``instructive mediation'' or ``evaluative mediation'') refers to how parents discuss the content with children during or after viewing; co-viewing refers to situations when parents and children watch television together, promoting parent-child connections \cite{valkenburg1999developing}. 

Apart from television, parent mediation theory has also been extended to explain the use of emerging technologies such as the internet \cite{durak2020parental}, video games \cite{nikken2006parental}, mobile devices/games \cite{sobel2017wasn}, creation-oriented educational media \cite{yu2021parental}, and conversational agents \cite{beneteau2020parenting}. First, \citet{nikken2006parental} suggested that parental mediation strategies in video-game play for children are comparable to those for television-watching, after conducting a survey-based study. They further explained how specific contextual factors such as parents' game behavior relate to these strategies. Moreover, parental mediation strategies were re-interpreted from traditional views for a location-based mobile game. Through a mixed-methods study, \citet{sobel2017wasn} found that the location-based mobile game, \textit{Pokemon Go} mitigated parents' traditional screen-time concerns, as it encouraged outdoor activity and promoted opportunities for parent-child bonding. This study exemplifies how the use of a different kind of technology can shift the interpretation of a restrictive mediation strategy (e.g., restriction of screen time) from a traditional perspective (e.g., great concerns about screen time) towards a modified view (e.g., mitigated concerns about screen time). Furthermore, \citet{beneteau2020parenting} recommended including the concept of augmented parenting in addition to the existing mediation strategies for a conversational agent. They investigated how voice-based interfaces, such as Alexa, affected parent-child dynamics at home and found that voice interfaces fostered communication between parents and children, disrupted each others' access to the technology through interruption or parental control, and augmented parenting practices. They argued that as the access to the technology is democratized (i.e., all family members have independent access to the device), parents might use technology to both mediate and augment parenting practices. Finally, \citet{yu2021parental} created a different mediation framework for the context of using creation-oriented educational media, inspired by the parent mediation theory. The newly-inspired mediation framework included creative mediation, preparation mediation, and administrative mediation. Each of these components partially reflected the notion of the three strategies from the original framework (i.e., restrictive mediation, active mediation, and co-viewing), but afforded its own meaning tailored to the specific educational context. 

The evolving framework or interpretation of parental mediation theory for different kinds of technology reflects the distinct needs of parental involvement, influenced not only by contextual factors but also by the affordances of the technology. However, very little has been done to explain the parental involvement needs for social robots in the context of education. In our work, we intend to explore how parents prefer and envision the adoption of a companion robot in their interactions with their child during in-home learning. 

Overall, studies related to interactions between a robot and a child or multiple children are relatively abundant. However, more attention needs to be oriented towards including adults in these interactions, given the importance of mediating technology use, especially  for young children. Although some work has been done to understand the interaction dynamic among a robot, a child, and a parent, questions related to how a parent would adopt a social robot to \textit{mediate the use of the robot} and  \textit{foster learning in children} as well as how \textit{the design for such interaction should be like} are yet unclear. We investigated a different age group from \citet{gvirsman2020patricc} and a different learning activity from \citet{chen2022designing}, while \citet{chan2017wakey} is not relevant to a learning context but daily routines. To fill in the research gap, we aim to address the following research questions in this work:

\begin{enumerate}
    \item \textit{RQ1: How do parents perceive and envision their use of a learning companion robot to facilitate a child's learning experience?}
    \item \textit{RQ2: What are the considerations for designing a learning companion robot for parents to engage their child in learning?}
\end{enumerate}

\begin{table*}[!bh]
    \caption{Participant Demographics}
    \label{tab:demographics}
    \centering
    \small
    \begin{tabular}{p{0.02\linewidth}p{0.08\linewidth}p{0.09\linewidth}p{0.09\linewidth}p{0.09\linewidth}p{0.19\linewidth}p{0.16\linewidth}}%
        \toprule
         \textbf{ID} & \textbf{Child Age (month)} & \textbf{Child \newline Gender} & \textbf{Parent \newline Gender} & \textbf{Ethnicity} & \textbf{Maternal \newline Education} & \textbf{Household \newline Income} \\
        \toprule
         1 & 55 m & Male & Female & White & Associate's (2-year college) & \$15,000-\$24,999\\
         \midrule
         2 & 60 m & Female & Female & Hispanic & High School & < \$15,000\\
         \midrule
         3 & 60 m & Male & Female & White & Master's & \$75,000-\$99,999\\
         \midrule
         4 & 55 m & Female & Female & White & Bachelor's & \$100,000-\$149,999\\
         \midrule
         5 & 44 m & Male & Female & White & Master's & \$150,000-\$199,999\\
         \midrule
         6 & 56 m & Female & Female & Biracial & Master's & \$100,000-\$149,999\\
         \midrule
         7 & 42 m & Male & Female & Black & Ph.D. & \$150,000-\$199,999\\
         \midrule
         8 & 53 m & Female & Female & White & Master's & \$50,000-\$74,999\\
         \midrule
         9 & 51 m & Male & Female & Biracial & Bachelor's & \$75,000-\$99,999\\
         \midrule
         10 & 42 m & Male & Female & White & Master's & \$75,000-\$99,999\\
         \bottomrule
    \end{tabular}
\end{table*}

\section{Method}

\subsection{Participants}
After obtaining institutional research ethics approval, we recruited families from cities in the Midwestern United States via email distributed to community centers and university employee mailing lists. We selected participants based on the following criteria: (1) one parent and one child were required to participate in the study together, (2) the child must have been between 3 and 5 years old, and (3) both the parent and the child were able to understand and communicate in English. We selected this specific age group as it corresponds to a period when informal learning experiences with parents at home are relatively important \cite{purpura2013informal}, occurring before children's active participation in formal schooling. Each family received \$25 USD as compensation upon completion of the study.

Our analysis includes data from 10 parent-child dyads, each with a child aged between 3 and 5 years old (4 female, 6 male, $M = 4.3$ years old, $SD = 0.58$ years old). Each study visit involved one parent and one child. Other parents and siblings were sometimes present nearby but did not take part in the study activities. Participant demographics are listed in Table \ref{tab:demographics}. We note that all parents who chose to participate in the study were female as mothers typically serve as primary caregivers {\cite{schoppe2013comparisons}} and are accordingly the ones who were willing to participate in the study. In addition, despite our efforts to recruit a diverse sample, our final sample offers limited socio-demographic representation (80\% of the mothers have at least a Bachelor's degree, 80\% of families have at least \$75,000 annual income), which may further constrain the generalizability of our results and omit insights into the unique needs of demographic groups that are less represented in our sample. We will discuss this limitation more thoroughly in Section {\ref{sec-5.3}}.


\begin{figure*}[!t]
  \includegraphics[width=5.8in]{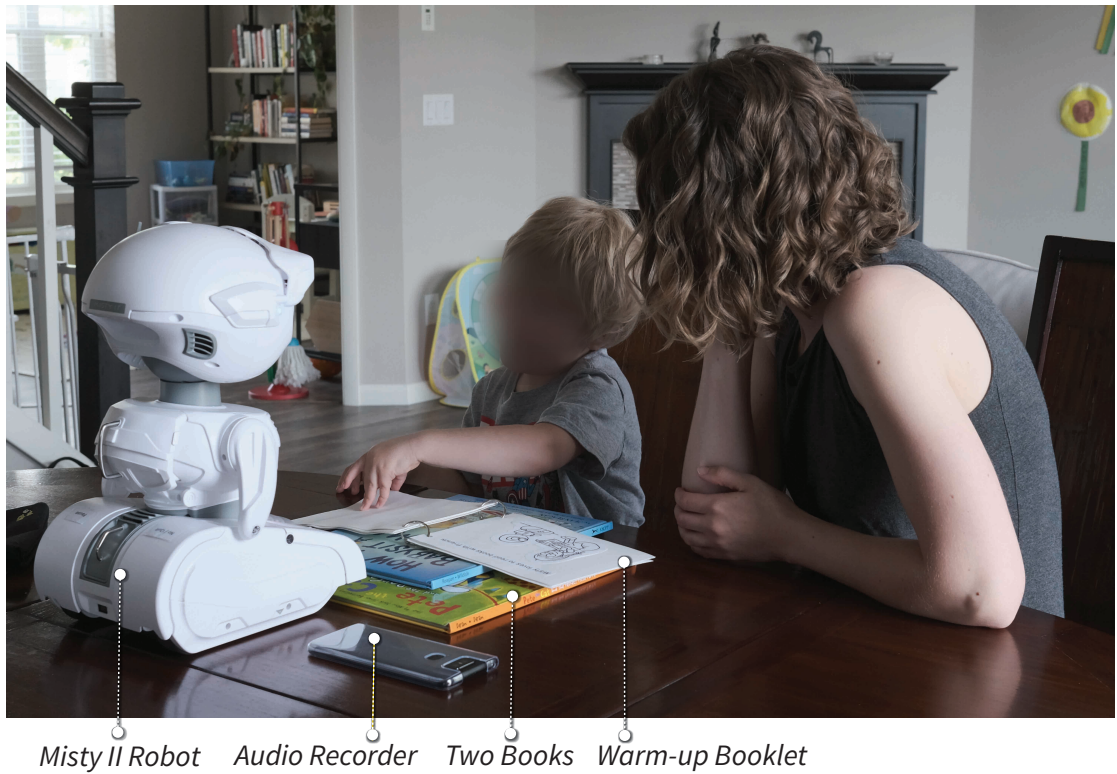}
   \vspace{-6pt}
  \caption{Our study materials included a social robot (Misty II), a warm-up booklet, two storybooks, and recording devices. Parent-child pairs read books at a table with the robot placed in front of them.}
  \label{fig:material}
   \vspace{-6pt}
\end{figure*}

\subsection{Study Design} 

We conducted short-term in-home visits with families and used a learning companion robot as a \textit{technology probe} to foster discussion on how a parent would engage and use the robot to support their child in learning and how they envisioned the robot should be designed for parent-child interaction. 

Our study objectives align with the three main goals described by \citet{hutchinson2003technology} about the technology probe approach--understanding users' needs and perception about the technology in a real-world setting, field-testing a technology, and motivating design of new technology. With the limited understanding of how parents might adopt a learning companion robot to interact with their child, introducing the learning companion robot with simple and flexible functionalities allowed us to explore unknown needs and preferences of families, understand their interpretation of how to use the technology, and inform new design opportunities for future improvements. In the following sections, we describe the details of the study materials and procedure. 

\subsubsection{Study Materials}

The study materials (Figure~\ref{fig:material}) included (1) a Misty II social robot,\footnote{Misty Robot: \url{https://www.mistyrobotics.com/products/misty-ii/}} which is a semi-humanoid robot with a 4-inch LCD display for its face where customizable facial expressions can be displayed; (2) a Raspberry Pi,\footnote{Raspberry Pi: \url{https://www.raspberrypi.org/}} which supports the software operations of the robot; (3) materials for the reading activity (i.e., one warm-up booklet and two story books)--the warm-up booklet was drawn and designed by the first author to familiarize parent-child dyads with how they can interact with the robot; two storybooks include one called \textit{Pete the Cat and the Perfect Pizza Party} (40 pages)\footnote{Book: Pete the Cat and the Perfect Pizza Party:\url{https://www.amazon.com/Pete-Cat-Perfect-Pizza-Party/dp/0062404377}} and the other called \textit{How to Babysit a Grandma} (32 pages),\footnote{Book: How to Babysit a Grandma: \url{https://www.amazon.com/How-Babysit-Grandma-relationships/dp/0385753845}} which were selected based on the appropriate reading age and book length; (4) recording devices (i.e., a video camera and an audio recorder), which were placed around the participants to capture participants' behavior and conversations.

\subsubsection{Study Procedure}

First, the experimenters offered the families an overview of the study and obtained informed consent: parents signed consent forms, and children gave verbal assents. Next, families were given the warm-up booklet to familiarize themselves with the learning companion robot. Then, families completed reading two storybooks with the robot. Finally, parents took part in a semi-structured interview on their experiences and reflections. Each study took around 70 minutes in total, with a 40-min reading session followed by a 30-min interview. Two experimenters were present throughout the reading session. Experimenters sat out of participants' sight to observe their interactions, take field notes, and answer technical questions from the participants.

\subsubsection{Interaction Design}

\begin{figure*}[t!]
  \includegraphics[width=\textwidth]{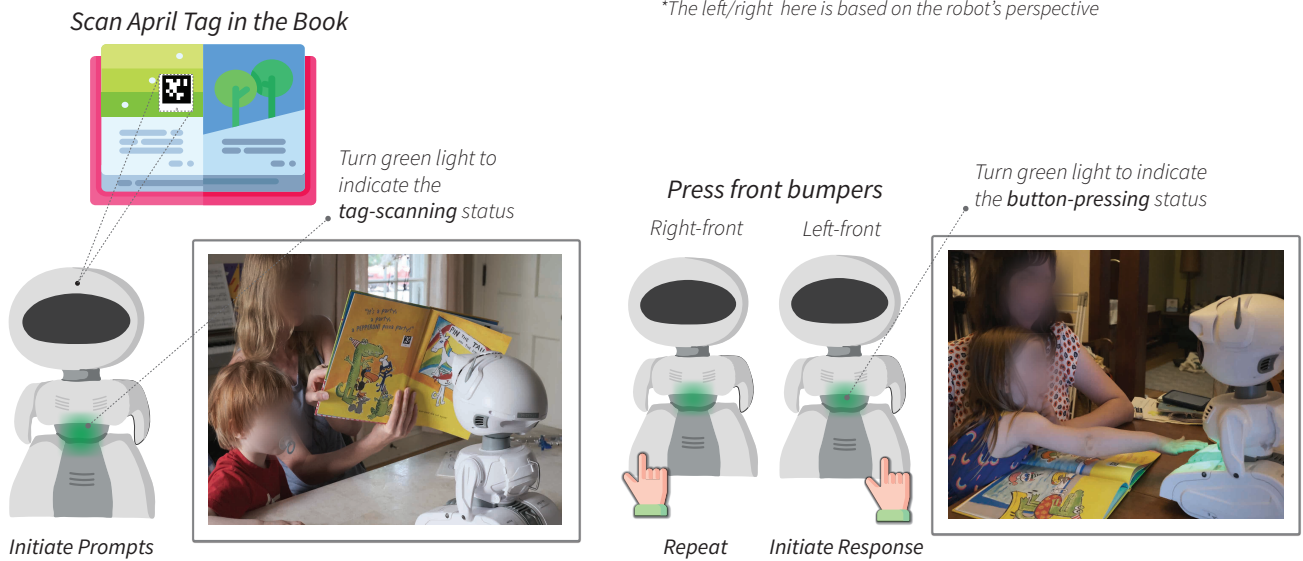}
   \vspace{-6pt}
  \caption{The interaction features of the robot: (A) users scanned tags in the book to initiate prompts, and (B) they pressed the right-front bumper to repeat what the robot said and the left-front bumper to initiate a response to the current prompt.}
  \label{fig:robotdesign}
   \vspace{-6pt}
\end{figure*}

\begin{figure*}[!t]
  \includegraphics[width=\textwidth]{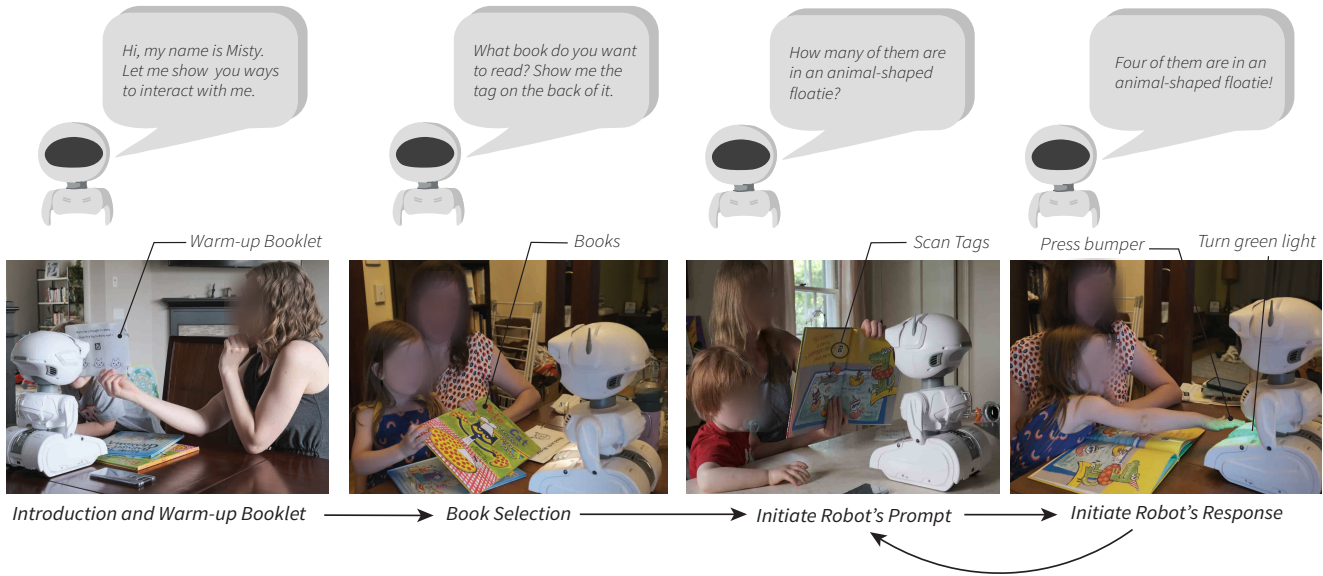}
   \vspace{-6pt}
  \caption{The interaction design for the reading activity. (1) The robot introduced itself to users and users used the warm-up booklet to learn and familiarize themselves with ways to interact with the robot. (2) Users selected a book to read. (3) Users initiated prompts from the robot when encountering a tag in the book during reading. (4) Users initiated a response to the current prompt from the robot by pressing the robot's left-front bumper.}
  \label{fig:design}
   \vspace{-6pt}
\end{figure*}

The main activity that the parent-child pairs did in the study was to read two books with the robot. In this section, we explain how users interacted with the robot during reading and how the prompts and responses of the robot were developed. The interaction design is also illustrated in Figure~\ref{fig:design}.

Before the families read the two books we provided,  they were given a \textit{warm-up booklet} which was drawn and designed by the first author in the format of a hand-made pictorial book, providing information and examples for how to interact with the robot. The booklet included guided activities that allowed families to practice showing tags to the robot, pressing bumpers to repeat what the robot previously said and obtaining responses. 

After the warm-up activity, families were told to finish reading two books, including \textit{Pete the Cat and the Perfect Pizza Party} and \textit{How to Babysit a Grandma}, with the robot and were given the choice to start with either one. Families were told that they were not required to initiate every tag and response, and were encouraged to use the features of the robot whenever they wanted to during the reading activity.

During the reading sessions, the robot provided \textit{verbal prompts} and \textit{responses}, generated using Google's cloud text-to-speech engine,\footnote{Google Cloud Text-to-Speech: \url{https://cloud.google.com/text-to-speech/}} when initiated by the participants. To initiate a verbal prompt, families needed to display April Tags\footnote{April Tags: \url{https://april.eecs.umich.edu/software/apriltag}} on the books to the robot. These April Tags were implemented in the book for every two pages and allowed the robot to autonomously recognize the given page and provide an appropriate prompt. To initiate a \textit{response} to the previously-initiated prompt, families needed to press the left-front bumper button along the robot's base. The \textit{Repeat} bumper on the right-front portion of the robot's base allowed families to repeat the robot's previous prompt. The robot's interaction features are also illustrated in Figure~\ref{fig:robotdesign}. We note that the experimenters did not instruct parent-child pairs regarding who should scan the tag or who should press the button. Parents and children interacted with the features naturally. It is possible that sometimes either the parent or the child always initiated the tags or the responses, sometimes both of them were involved interchangeably, and sometimes they never press the button to obtain the robot's responses as the parents were confident about the answers.

The reading activity is based on the notion of \textit{math talk}, where parents incorporate math concepts in daily-life conversations with their child to foster development of math concepts. We developed prompts related to basic and advanced math concepts for preschoolers based on literature in educational psychology \cite{purpura2013informal, engel2013teaching} and these prompts were delivered by the robot when participants show the corresponding tags on the book to the robot. The reading activity allowed parents to engage in problem solving and math-oriented conversations with their child. We chose reading as it is a common parent-child shared activity and has also been widely used to study in previous work to support language and math learning \cite{vandermaas2009numeracy, purpura2017causal}. 

The framework underlying the design of prompts implemented in the robot was developed based on the informal numeracy structure proposed by \citet{purpura2013informal} and the classification rules of math proficiency levels proposed by \citet{engel2013teaching} for preschool children. Specifically, the informal numeracy skills \cite{purpura2013informal} were categorized into one of the four math proficiency levels specified in the framework proposed by \citet{engel2013teaching}.

\subsubsection{Semi-structured Interview}
We conducted a semi-structured interview with parents after they finished reading two storybooks with the child while engaging the learning companion robot. The interview protocol included questions about parents' overall experiences (e.g., ``\textit{What do you think about the interaction with the robot overall}''), comparisons with their regular informal learning experiences at home (e.g., ``\textit{How would you compare this reading experience with your regular ones at home}''), opinions on what they preferred the robot to support (e.g., ``\textit{How would you prefer the robot to support you}''), and how they envisioned the use of the robot in their daily scenarios related to children's learning (e.g., ``\textit{How would you see yourself using the robot in your daily life to support [child]'s learning}''). Other questions were generated by the first author during the study, based on the parents' responses to the open-ended questions from the protocol and the field notes taken during their reading sessions. Interview sessions were audio recorded and transcribed for analysis.

\subsection{Data Analysis}

To anonymize our participants in the data, we report the findings using participants' ID assigned by the experimenters based on their order of completing the study. For example, P1 refers to the first participant. We conducted a reflexive \textit{Thematic Analysis (TA)} based on \citet{clarke2014thematic} and \citet{mcdonald2019reliability} on the semi-structured interviews to identify patterns in the data and derive a grounded understanding of parents' perspectives. First, the first author and an undergraduate assistant, who is experienced in qualitative coding, transcribed and familiarized themselves with the data based on the audio and video recordings. Second, the first author generated initial semantic codes based on her interpretation of the data and categorized the data based on the codes. For example, code ``leverage robot's responses to engage children'' and code ``use robot's answers as a source of validation'' were both categorized as ``parents' use of robot's response.'' Third, both the first author and the undergraduate assistant coded the data from P1 with the initial codes, followed by an iterative discussion to find consensus on interpretations and emergent candidate themes and to generate an agreed code book. Fourth, for the rest of the data, the first author and the undergraduate assistant used the code book to code the rest of the data independently and peer-reviewed each others' codes. For instance, the first author coded P2 and the assistant peer-reviewed P2; the assistant coded P3 and the first author peer-reviewed P3. The finalized codes were reviewed and discussed iteratively. Lastly, we reviewed, refined, and constructed the themes as findings based on meaningful and insightful patterns.

\begin{figure*}[t!]
  \includegraphics[width=\textwidth]{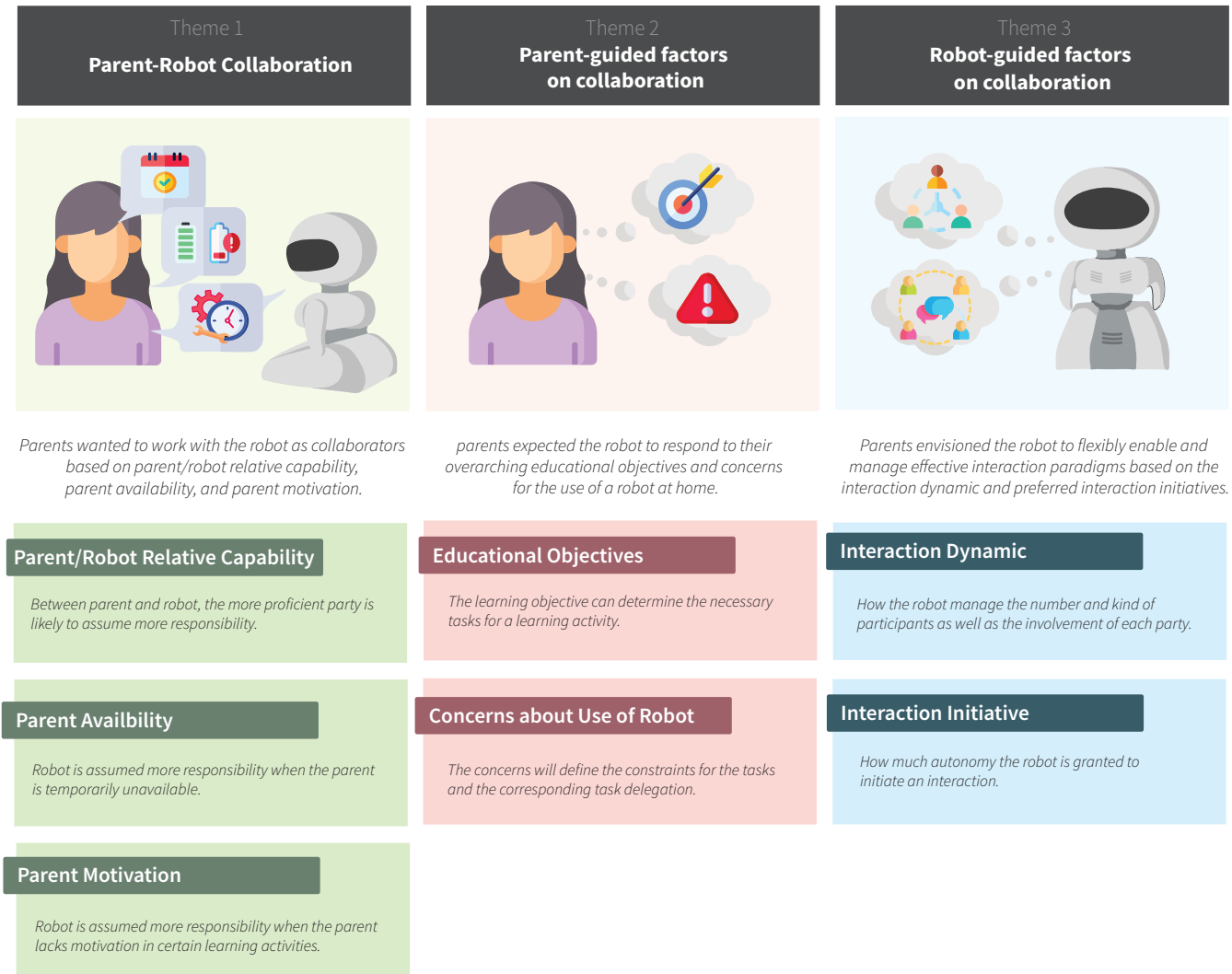}
   \vspace{-6pt}
  \caption{An overview of the results. We identified three main themes, including (1) parent-robot collaboration (parent/robot relative capability, parent availability, and parent motivation), (2) parent-guided factors (educational objectives and concerns about use of robot), and (3) robot-guided factors (interaction dynamic and interaction initiative).}
  \label{fig:overview}
   \vspace{-6pt}
\end{figure*}

\section{Results}

Our analysis of interview data from parents highlights three interacting themes informing the design of a learning companion robot that supports the context-based needs for parent-child interaction. The first theme, \textit{parent-robot collaboration}, highlights parents' wanted to work with the robot as \textit{collaborators} based on relative capabilities between the parent and the robot, the availability of the parent, and the motivation of the parent. A consideration of the \textit{relative capabilities} of the parent and the robot would involve assigning more responsibility in a learning task to the party with higher skill on that task \cite{horvitz1999principles}. For example, the task of generating educational prompts might be delegated to the robot, while the parent might offer personalized scaffolding for children. The parent's \textit{availability} can help determine times when the parent has limited availability, such as when taking care of younger siblings, and assign more responsibility to the robot. Finally, the parent might situationally lack \textit{motivation} in pursuing some learning activities or repeating prompts, which might provide an opportunity for the robot to assume more responsibility.

The second theme, \textit{parent-guided factors on collaboration,} underlines parent expectations for the robot to respond to their overarching \textit{educational objectives} and \textit{concerns} for the use of a robot at home. The learning objective can determine the necessary tasks for a learning activity. For example, the parent's learning objective might be understanding the process of problem solving, in which case either the parent or the robot needs to explain the process rather than only provide an answer. The concerns will define the constraints for the tasks and the corresponding delegation of the tasks, unless a design feature effectively mitigates the concern. For example, the parent's concern might be about the privacy of their video data, in which case they may not delegate tasks that require computer vision or video capturing to the robot. And the third theme, \textit{robot-guided factors on collaboration,} highlights parents' vision that the robot flexibly enables and manages effective interaction paradigms, including the concept of \textit{interaction dynamics} and \textit{interaction initiatives}, in line with the collaboration patterns and parental expectations. The notion of interaction dynamic refers to how the robot manages the number and kind of participants as well as the involvement of each party. For example, the robot might need to switch to a different interaction mode when the parent is involved in its interaction with the child. The idea of interaction initiatives pertains to how much autonomy the robot is granted to initiate an interaction. For example, the robot might initiate an interaction anytime when free playing but initiate only as requested when reading with the child. We present the details of each theme in the following sections, and a summary of these findings can be found in Figure~\ref{fig:overview}. 

\begin{figure*}[h!]
  \includegraphics[width=\textwidth]{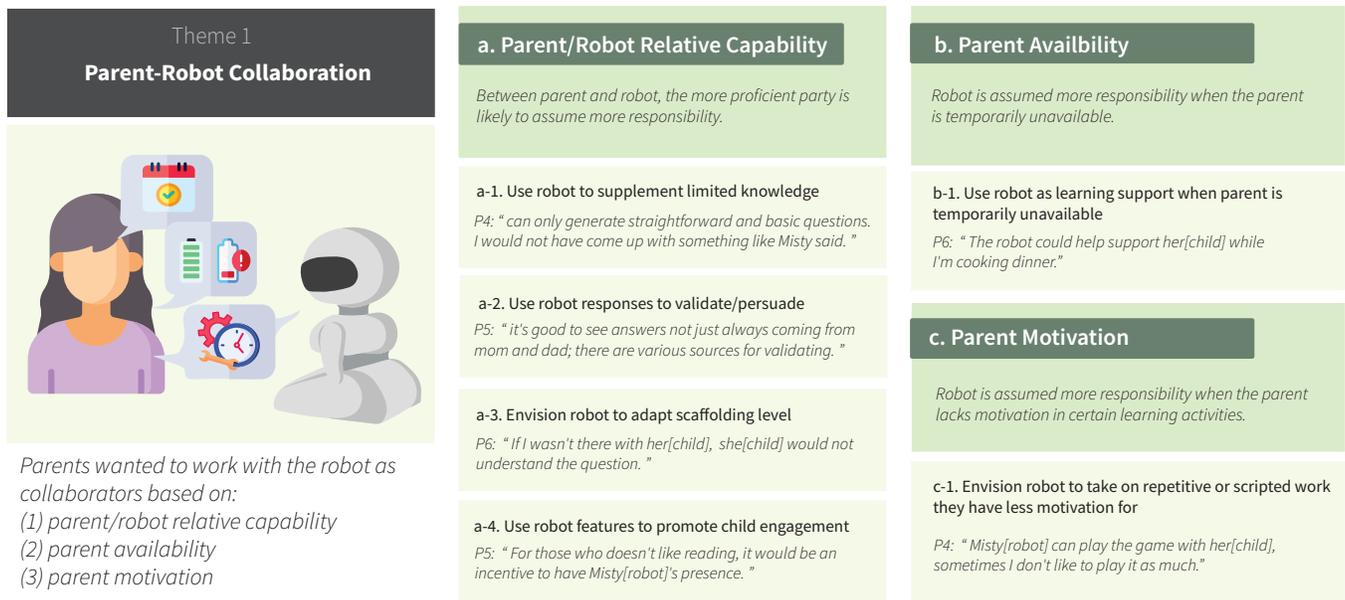}
  \vspace{-6pt}
  \caption{Summary of Theme 1: Parent-robot collaboration. We found that parents wanted to work with the robot as collaborators based on parent/robot relative capability (C), parents' availability (A), and parents' motivation (M).}
  \label{fig:theme1}
   \vspace{-5pt}
\end{figure*}

\subsection{Parent-Robot Collaboration}\label{sec-4.1}

In our data, parents frequently discussed several tasks (e.g., generating creative educational prompts for specific learning concepts, offering personalized scaffolding, etc.) that must be completed to achieve the learning goals they expected for their child. More specifically, they illustrated how these tasks should be delegated between themselves and the robot as they reflected on their as well as the robot's strengths and weaknesses in daily learning scenarios. This pattern in our data suggested that parents were thinking more about how they could collaborate with the robot, which was further verified by their explicit description for the robot's role as their ``\textit{collaborator}.'' In addition, task delegation choices were based on three dimensions: parent/robot relative capability, parent availability, and parent motivation. A summary of these findings can be found in Figure~\ref{fig:theme1}.

\subsubsection{Robot as a collaborator}  The majority of parents (P3, P4, P5, P7, P8, P9, P10) wanted the robot to act as their collaborator to alleviate some of their workload rather than having the robot replace their responsibility entirely. P5 summarized this perspective by stating, ``\textit{I think the robot is partnering with me. I don't think it's taking the place of me.}'' In addition, P8 highlighted the robot's role as a team member to alleviate her authoritative stance with her child, saying ``\textit{Misty and I are a team and we are working together to try to grow those concepts.}'' Furthermore, P9 and P10 emphasized how the robot offloaded pressure from them by generating appropriate intellectual prompts during their interactions. As articulated by P9, ``\textit{parent is still very involved in being the teacher, but the robot takes the pressure off of the parent to have to come up with questions.}'' Although some parents also used different terms to describe the robot's functional role as, for instance, a facilitator (P4, P7), a caregiver (P1, P2), or an educator (P1, P7), their articulations of how the robot served these roles are still conceptualized as a \textit{collaborator}. We argue that parents perceived the robot's functional roles differently depending on which tasks the robot collaborated with parents on. For example, the robot may have been perceived more as a facilitator if it helped with prompting parent-child conversations. On the other hand, the robot may have been perceived more as a tutor if it helped with demonstrating a problem solving process. In whichever case, the robot served as a \textit{collaborator} to the parent by offering different kinds of support, leading to the perception of different functional roles while staying under the umbrella of a \textit{collaborator}. In the following sections, we further describe what factors should be considered to determine the complicated patterns of parent-robot collaboration.

\subsubsection{Parent/Robot Relative Capability} The relative capability between the parent and the robot refers to the strengths and weaknesses of the parent and the robot with respect to each other in carrying out certain tasks. For each task, the more proficient party is likely to assume more authority. We describe the collaborative tasks that emerged from our data through the lens of this perspective in the following sections.

\paragraph{\textbf{Parents can use the robot to supplement their limited knowledge domains for prompt generation.}} All parents (P1--P10) mentioned that spontaneously generating prompts associated with advanced or unfamiliar concepts might have been beyond their capabilities. Despite their comprehension of the underlying academic concepts, they recognized their limited knowledge of the pedagogical strategies required to independently create age-appropriate prompts. For instance, P2 admitted that she would find herself ``\textit{reliant on the content provided in the book},'' while P4 mentioned that she could only generate ``\textit{straightforward and basic questions},'' lacking the ability to formulate more advanced ideas. Additionally, P10 perceived the collaboration to be based on hers and the robot's respective proficient domain knowledge, stating ``\textit{I read the story and provide literacy-related interaction, and the robot can assist by suggesting math-related questions for them[children] to think about.}'' Moreover, parents not only sought academic support but also envisioned prompting the child to reflect on social interaction and emotions, nurturing their emotional intelligence. For example, P3 and P5 believed that the robot could play a pivotal role in generating prompts encouraging children to reflect on their own and others' feelings and emotions. These findings suggest that in cases where parents lacked specific knowledge, the robot could assume the role of generating advanced concepts and prompts to facilitate children's learning.
  
\paragraph{\textbf{Parents can use the robot's responses as a source of validation and social persuasion method.}}
The majority of parents (P1, P4, P5, P6, P7, P8) leaned towards leveraging the robot's responses as a source of validation. They saw the responses as a means to offer their children contextualized feedback and encouragement. P8 exemplified this concept, stating, ``\textit{one thing I liked is that after [child] came up with the wrong answer, and misty[robot] told us the right answer. And then we could work backwards and figure it out together.}'' Interestingly, several parents (P4, P5, P7, P8) explained that they utilized the robot's response because their children trusted third-party and non-parental opinions more than their input. P4 summarized this phenomenon by noting, ``\textit{if they have feedback from someone who's other than a parent, they seem to take that feedback better.}'' One possible rationale, articulated by P5, is that ``\textit{it's good to see answers not just always coming from mom and dad; there are various information sources validating that. } Another explanation, as described by P8, is that she thought her child had ``\textit{never seen Misty made a mistake, unlike her parents, so she has more chance to trust Misty's abilities.}'' However, in the spectrum of parental preferences regarding the robot's responses, extreme cases exist, where parents either expressed a desperate need (P2) or no need at all (P3) for receiving a response or an answer to a prompt from the robot. P2 admitted feeling uncertain about offering a proper response for her child, while P3 demonstrated stronger leadership in guiding her child in line with her parenting values, stating that ``\textit{It's not really about getting the right answer. It's more about the interaction between us and having open conversations about the questions.}'' These findings indicate that the robot could act as a source of validation for facilitating learning interactions, be leveraged as a trusted third-party opinion to effectively persuade and guide children, and support parents who lacked confidence in their own responses.
  
\paragraph{\textbf{Parents envision the robot tailoring the level of scaffolding to their capabilities.}}
In the current system design, many parents (P1, P4, P5, P6, P8, P10) expressed that they needed to provide substantial scaffolding during the interaction with the robot. This scaffolding encompassed activities such as clarifying the meaning of questions (P4, P6, P8) and explaining the problem-solving process (P1, P4, P5, P6, P8, P10). For instance, P6 stated that ``\textit{If I wasn't there with her[child],  she[child] would not understand the question, although She[child] understood the math part of it.}; P1 noted, ``\textit{Robot didn't explain, but for young kids, they wouldn't necessarily know unless someone is explaining to them.}'' However, one parent (P2) encountered challenges in providing scaffolding for her child. She expressed uncertainty, stating, ``\textit{I am unsure what to do}, after observing that her child answered incorrectly. Unlike most other parents, she pressed the robot's bumper button to obtain the standard responses without providing her child any feedback. These findings highlight the need for the robot to tailor its level of scaffolding to the parents' capabilities to optimize the learning experience.  
  
\paragraph{\textbf{\textit{Robot could enhance child's engagement with physical and interactive features}}}
Parents excelled at customizing their approaches based on their understanding of their child's interests (P1, P7, P10).  For instance, P7 knew how to``\textit{use something he[child] is interested in get their attention.}'' Simultaneously, parents recognized the significance of the robot's physical appearance (P3, P4, P5, P8, P9) and interaction features such as buttons (P3, P6, P8, P9, P10) in maintaining their child's engagement and motivation. As P5 described, for someone ``\textit{who doesn't like reading, it would be an incentive to have Misty[robot]'s presence.}'' Moreover, when asked to compare delivering prompts with the robot through speech versus via text, P3 mentioned that she would ``\textit{be more interested in robot than paper because it would engage her child better.}'' P8 also emphasized the robot's interactive feature, saying, ``\textit{she really liked to push the button so that kind of kept her engaged.}'' In summary, these findings indicate that while the robot may not excel at offering customized motivational approaches as effectively as parents, its physical and interactive features can assist parents in enhancing their child's engagement.

\subsubsection{Parent Availability}
The parent's availability dimension helps identify times when the parent has limited availability to support the child. Although we did not discuss the robot's availability respectively as it did not emerge from the discussion, it could contextually mean the battery level of the robot or the ease of access to the robot.

\paragraph{ \textbf{Parents can use the robot as educational support when they are temporarily unavailable.}} Parents (P1, P4, P5, P6, P8, P9) recognized that there are occasions when they cannot be available to assist their child. These scenarios may occur due to various commitments such as household chores (P6), work (P4, P6), caring for other siblings (P5), or even during the recovery of an unexpected incident (P1). P5 described this situation by saying, ``\textit{I have two other kids, so sometimes I'm occupied with helping them, and I can't be there.}'' Similarly, P6 suggested that ``\textit{the robot could help her[child] for five minutes while I'm cooking dinner.}'' Furthermore, P1 shared a personal experience, stating, ``\textit{Last year, I was in a severe accident and bedridden for three to four months. During that challenging period, I might have taken the assistance of a robot to entertain my kids when I was physically unable to do so.}'' Additionally, P9 highlighted that she preferred the robot to support her child doing learning activities instead of being occupied by entertainments, stating, ``\textit{When we are busy, we would feel better about kids engaging in educational activities with the robot than watching TV.}'' These parents' perspectives emphasize the potential role of the robot in offering educational support when parents are temporarily unavailable or occupied with other responsibilities.

\subsubsection{Parent Motivation}

The parent's motivation dimension helps identify times when parents lack motivation in certain learning activities or tasks. We did not discuss the robot's motivation respectively as it did not emerge from the discussion and we also assumed this dimension is not applicable to the robot.

\paragraph{\textbf{Parents envisioned the robot taking on repetitive and scripted work that they have less motivation for.}} Several parents (P1, P2, P3, P4, P6, P9) expressed a preference for the robot to handle certain tasks, particularly those that are repetitive, structured, and scripted. For example, some parents (P1, P2, P3, P6) showed the desire to have the robot read the content of the books, allowing them to focus on providing personalized and instructional support. P6 illustrated this by stating, ``\textit{Misty[robot] can read the book to her[child], but I don't think that she[child] would understand additional instructions. I need to explain to her.}'' Furthermore, P3 wanted the robot repeat stories to her child, so he can ``\textit{hear it again in a different way from somebody else.}'' P4 extended this idea to educational games, as her child enjoys playing them repeatedly. She envisioned having ``\textit{Misty[robot] play the game with her[child]}'', reducing her involvement in repetitive tasks as she ```\textit{sometimes don't like to play it as much.}'' This illustrates parents' intention to delegate lower-level and repetitive responsibilities to the robot, freeing them to engage in more interactive and personalized interactions with their child.

\begin{figure*}[h!]
  \includegraphics[width=\textwidth]{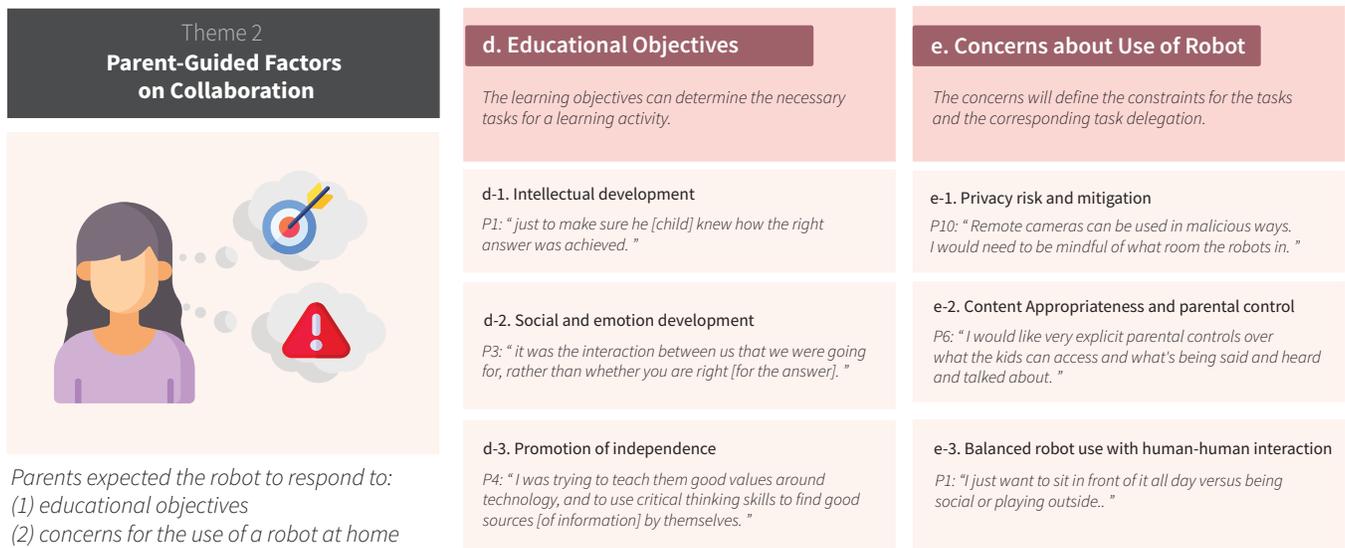}
   \vspace{-6pt}
  \caption{Summary of Theme 2: Parent-guided factors on collaboration. We found that parents expected the robot to respond to their overarching educational objectives (intellectual development, social and emotion development, and promotion of independence) and concerns for the use of a robot at home (privacy, content appropriateness, and balance between human-human and human-robot interactions.)}
  \label{fig:theme2}
   \vspace{-6pt}
\end{figure*}

\subsection{Parent-Guided Factors on Collaboration}\label{sec-4.2}

Parents formulated their individual policies for using the social robot, while upholding their core educational objectives when engaging in learning activities with their child. Even when guided by prompts from the social robot, they instinctively tailored their parental guidance to fill in any gaps and ensure the attainment of these objectives. This not only reveals the parents' aspirations in supporting their children but also highlights potential design implications for the social robot. We categorized these learning objectives into three domains: intellectual development, social and emotion development, and promotion of independence. A summary of these findings can be found in Figure~\ref{fig:theme2}.

\subsubsection{Educational Objectives}

\paragraph{\textbf{\textit{Intellectual development}}} For the majority of parents (P1, P2, P4, P5, P6, P8, P10) their primary goal was to nurture their children's comprehension of the problem-solving process, prioritizing it over the mere attainment of correct answers. As articulated by P1, her goal was ``\textit{just to make sure he [child] knew how the right answer was achieved},'' and P10 added that she was ``\textit{not super concerned with him getting the answer right.}'' However, P2 noted that she was ``\textit{unsure or didn't know what to do}'' when her child encountered difficulties in understanding how to tackle a problem, highlighting the challenges she faced in providing effective guidance to achieve her desired educational outcomes. Additionally, beyond the realm of problem-solving, P2 mentioned her efforts to teach her child Spanish and English to facilitate bilingual communication, allowing her child to interact effectively with both family members at home and peers at school: `` \textit{I tried to speak more Spanish at home, because they speak English in school. Their grandma doesn't speak English, so we have to speak Spanish with her.}''
  
\paragraph{\textit{\textbf{Social and emotion development}}} Some parents expressed that one of the objectives in doing learning activities at home is to spend time together to strengthen their parent-child bond (P3, P9, P10) and promote the child's comprehension of social interactions with others, including the ability to recognize and empathize with others' emotions (P3, P5, P7, P9). For example, P3 mentioned that ``\textit{it was the interaction between us that we were going for, rather than whether you are right [for the answer].}'' In addition, P9 described her aim saying: ``\textit{we're more focused on social and emotional learning, and get him [child] to talk about feelings.}''

\paragraph{\textit{\textbf{Promotion of independence}}} Parents believed that a substantial aspect of learning involves cultivating the independence of acquiring knowledge (P3, P1, P5, P7) and making critical judgements on information from various sources (P4). As articulated by P1, her aim was to ``\textit{try to get him to figure things out on his own.}'' Similarly, P4 shared that she was ``\textit{trying to teach them good values around technology, and to use critical thinking skills to find good sources [of information] by themselves.}'' In addition, this perspective was cited as a rationale for favoring occasional one-on-one child-robot interactions, a choice contingent upon the specific learning goals at any given moment (P5). As specified by P5, she thought ``\textit{there's some value in interacting and stepping back to let them process things for themselves and try to understand stuff, which also depends on what the goals are in that moment.}''

\subsubsection{Concerns and mitigation approach}
Parents expressed concerns about using the social robot at home and discussed potential solutions. These findings shed light on the factors that limit the robot's usage and offer insights into how design improvements could help alleviate parental concerns. Three primary concerns emerged from our analysis: privacy, content appropriateness, and achieving a balanced use of the robot.

\paragraph{\textbf{Privacy risk and mitigation}}
While some parents voiced apprehensions regarding privacy, with a particular focus on data security (P1, P4, P6, P8), and their child's understanding of data privacy (P4), others (P9, P10) explicitly stated that they had no such concerns. Specifically, P8 harbored concerns about the robot's installed cameras potentially capturing in-home video data, cautioning that ``\textit{Remote cameras can be used in malicious ways. I would need to be mindful of what room the robots in.}'' Meanwhile, P4 voiced uncertainty about the content and timing of data recording, raising questions such as ``\textit{Is it always quietly listening, waiting for a voice activation feature? Can I actively turn it on and off?Is it recording my house all the time?}'' Additionally, both P4 and P6 shared concerns about their child inadvertently disclosing personal information to the robot without a comprehensive understanding of how data would be handled, including unexpected sharing and retention. P4 emphasized this by remarking ``\textit{even just reminding my kids that what goes on the internet really is forever.}'' 

Parents offered several suggestions to mitigate their privacy concerns, emphasizing transparency (P4), clear explanation (P4), and reduced reliance on internet connectivity (P8). P4 proposed that data recording, storage and usage should be transparently and comprehensibly explained through the system interface in a user-friendly manner. She said, ``\textit{I would want transparency about what the robot is doing. If it is collecting data. I want to know what it's collecting.}'' Meanwhile, P8 expressed a preference for the robot's operation not to be contingent on an internet connection to safeguard data privacy, stating ``\textit{If Misty could download the stories and the question so it will not be Wi-Fi dependent, I would feel more comfortable with it.}''
  
\paragraph{\textbf{Content Appropriateness and parental control}}
Several parents raised concerns about the age-appropriateness (P2, P4, P5, P6, P7, P10) and trustworthiness (P6) of the content delivered by the robot. They shared the desire to prevent the robot from presenting questions or materials they considered inappropriate. Parents held varying perspectives on how best to control and filter the content. The majority strongly preferred to personally evaluate and determine content suitability (P4, P5, P6, P10). As remarked by P4, ``\textit{I would have the ability to say--Yes, this is appropriate; No, this is not.}'' Additionally, P6 added that she wants ``\textit{very explicit parental controls over what the kids can access and what's being said and heard and talked about.}'' Aside from simply filtering the content, P10 emphasized her desire to be ``\textit{involved in what the child is doing, instead of giving him the robot and leaving.}'' This hands-on approach allowed her to actively supervise her child's usage in line with her parental expectations. It's worth noting that these observations stem from a short-term study and may evolve as parents establish increased trust in the robot over time. Further elaboration on this point is provided in Section~\ref{sec-5.3}. 

\paragraph{\textbf{Balanced robot use with human-human interaction}}
Several parents (P1, P5, P6, P7, P9) expressed concerns about their child potentially overusing the robot and consequently neglecting other important aspects of learning and development, such as outdoor activities and social interaction. P1 worried that her child might become excessively absorbed in electronic devices and ``\textit{just want to sit in front of it all day versus being social or playing outside.}'' Similarly, P6 expressed concerns about children becoming isolated, primarily engaging with technology instead of learning how to interact with others: ``\textit{I feel like kids are not learning how to interact with each other and just want to play by themselves, with technology involved.}'' Despite these concerns, parents recognized the value of the robot in enriching their children's learning. P7 voiced a preference for technology to improve their lives positively rather than exerting excessive control: ``\textit{It's more like I want technology to enhance their lives instead of letting it control them.}'' This sentiment highlights the parental aspiration for a balanced integration of technology into their child's development.

\begin{figure*}[h!]
  \includegraphics[width=\textwidth]{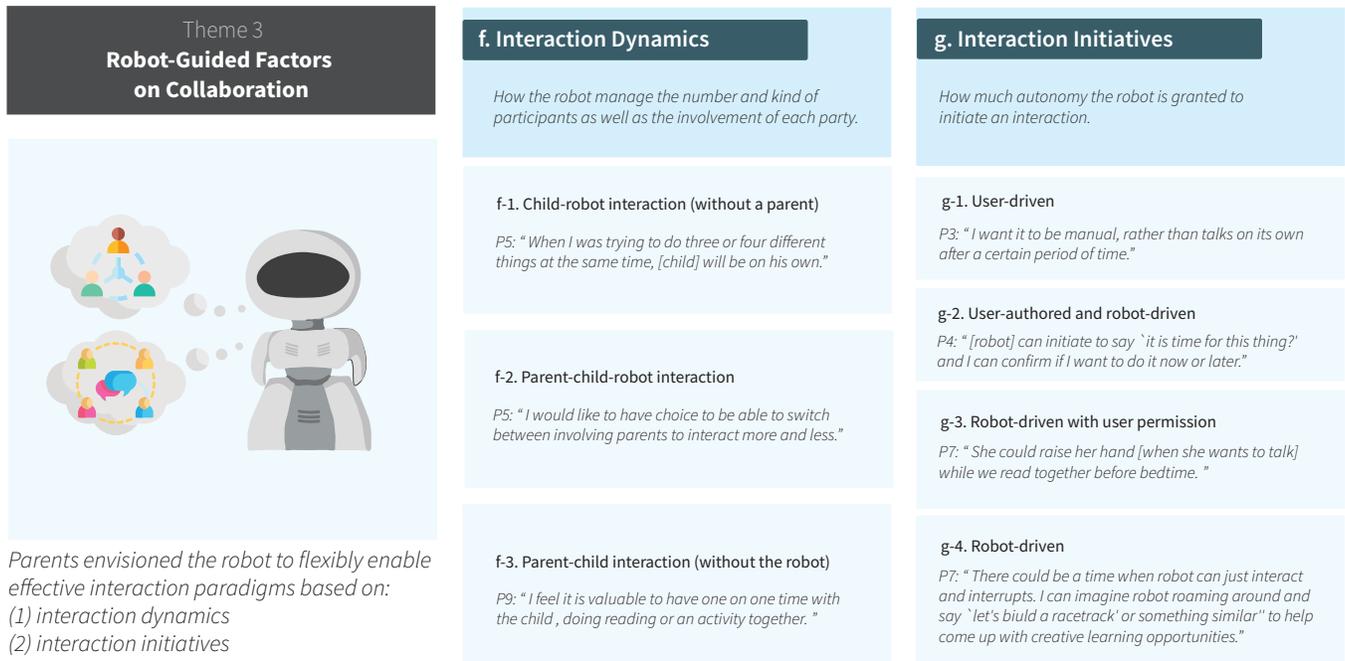}
   \vspace{-8pt}
  \caption{Summary of Theme 3: robot-guided factors. We found that parents envisioned the robot enabling effective interaction paradigms by handling various interaction dynamics (child-robot interaction, parent-child-robot interaction, parent-child interaction) and mixed-initiative forms of interaction (user-driven, user-authored and robot-driven, robot-driven with user permission, and robot-driven).}
  \label{fig:theme3}
   \vspace{-6pt}
\end{figure*}

\subsection{Robot-Guided Factors on Collaboration}\label{sec-4.3}

The concept of \textit{interaction paradigm} defines a specific module of interaction with the robot, governed by the interaction dynamic and the preferred initiation of communication. In our data, we found that parents preferred the robot to flexibly supplement their level of involvement and they favored taking more control over when the robot initiates an interaction. A summary of these findings can be found in Figure~\ref{fig:theme3}.

\paragraph{\textbf{Parents wanted the robot to adapt to flexible choices of interaction dynamics.}} The concept of interaction dynamics refers to how the robot manages the number and kinds of participants as well as the involvement of each party. Instead of adhering to a fixed mode, most parents (P4, P5, P6, P8, P9, P10) wanted the robot to flexibly support a range of interaction dynamics based on situational needs. For instance, P5 illustrated that she would like to ``\textit{have choice to switch between having parents in the interaction or letting the child be on his own.}'' Parents conveyed a variety of dynamics they preferred in different scenarios with the robot, including child-robot interaction without a parent, parent-child-robot interaction, and parent-child interaction without the robot. 

Firstly, in the dynamic of \textit{Child-robot interaction (without a parent)}, parents leaned towards allowing their child to interact with the robot independently when they are temporarily unavailable (P4, P5, P6, P8, P9), when they lack motivation (P1, P9), as the child matures to an older age (P2, P5), or when the objective is to nurture their child's independence for self-directed learning (P5, P8). For instance, P5 described that when she was ``\textit{trying to do three or four different things at the same time, [child] will be on his own.}'' Moreover, P9 indicated that she prefer to do more outdoor activities ``\textit{rather than learning activities at home with her child,}'' and thus would favor a child-robot interaction model for in-home learning activities. In addition, P2 and P5 both imagined their children gradually interacting with the robot on their own as they get older. Lastly, P8 illustrated her desire to promote independent learning for her child, stating that the robot ``\textit{provide the response that gives her the tools to be a little bit independent and practicing the skills without me''} (P8).

Secondly, in the dynamic of \textit{Parent-child-robot interaction}, all parents (P1---P10) expressed their preference for a flexible division of roles in guiding interactions with the child, allowing either the robot or themselves to take the lead while the other played a supplementary role. This adaptable approach results in varying levels of involvement for each party and thereby variations within this specific dynamic. Firstly, parents leaned towards increased robot involvement when they lack the availability (P5, P6), motivation (P4), or capability (P2, P8) to actively guide the activities, when they aim to cultivate their child's independence in learning (P5). For example, P8 described the lack of experience her husband has in facilitating children's learning, stating that ``\textit{my husband does not know how to facilitate and he wouldn't enjoy getting emails about it. The robot could guide his interaction and maybe have it model him[husband] more.}'' Conversely, parents themselves wanted more active involvement when they seek quality bonding time with their child during learning (P4, P6), when they intend to contribute specific intellectual concepts (P4, P10), when personalized support is deemed necessary, or when they want to maintain supervisory control over their child's interaction with the robot to address any concerns (P7). P10 illustrated this when she stated ``\textit{I think it is always beneficial from having some parental interaction and interest in what he's doing. So I want to be there to encourage him and let him validate that he's doing the job.}'' Combining these two views, parents prefer to have choice in both. P5 explained that she ``\textit{would like to have choice}'' in the dynamics, and be able to ``\textit{switch between involving parents to interact more and less.}'' In particular, she stated that ``\textit{I can provide guidance or assistance, but only being pulled in when needed, letting him do what he needs to do by himself.}''

Finally, in the dynamic of \textit{Parent-child interaction (without the robot)}, a few parents preferred to interact with their child without the robot when they want to enjoy their own time together (P9), when the concepts they want to explain are not enabled by the robot (P8), or when they have concerns over the information the robot offers (P6). For instance, P9 illustrated the situation when she said ``\textit{I feel it is valuable to have one on one time with the child, doing reading or an activity together.}''

\paragraph{\textbf{Parents envisioned the robot accommodating flexible forms 
of interaction initiatives.}} The notion of interaction initiative refers to how much autonomy the robot is granted to initiate interactions with users. Our data suggested different patterns of initiating an interaction including user-driven, user-authored and robot-driven, robot-driven with user permission, and robot-driven. In the following sections, we articulate each kind of initiative and the the possible scenarios correspondingly. 

First, in a \textit{user-driven} case, parents (P2, P3, P7, P8, P10) preferred the robot to initiate an interaction only when they manually request one. P3 described that she preferred the robot ``\textit{to be manual, rather than talks on its own after a certain period of time.}'' Second, in a \textit{user-authored and robot-driven} case, where parents prefer to author the timing and initiatives of the robot. P4 described that she ``\textit{wouldn't want the robot to just be able to barge in to say whatever it wants or ask for permission randomly.}'' She further illustrated her intention to author the robot's initiatives by scheduling a specific prompt or behavior for a specific time frame, stating, ``\textit{it[robot] can initiate to say, `it is time for this thing, do you want to do it now,' and I can confirm if I want to do it now or later.}'' Third, in the case of \textit{robot-driven with user permission}, parents (P2, P7) described certain scenarios when the robot can drive the initiative by indicating its intention but need to wait for user permission before actually putting the initiative into actions. P7 described that she preferred the robot to ``\textit{raise her hand while we read together before bedtime}'' to keep the interaction ``\textit{at a lower intensity level.}'' Finally, in a \textit{robot-driven} case, parents (P6, P7) described that sometimes the robot can initiate an interaction or conversation anytime. P7 exemplified this case by describing a scenario when her child is playing toys: ``\textit{There could be a time when she can just interact and interrupts. I can imagine her roaming around and say `let's build a racetrack' or something similar'' to help come up with creative learning opportunities.}''
\section{Discussion}

Given the importance and ignorance of parental involvement in children's adoption of learning companion robots, we aimed to answer the following research questions: (1) How do parents perceive and envision their use of a learning companion robot to facilitate a child's learning experience? (below and Section~\ref{sec-5.1}); (2) What are the considerations for designing a learning companion robot for parents to engage their child's learning? (Sections~\ref{sec-5.2} and \ref{sec-5.3})

Through a technology probe study with contextual interviews, we found that most parents wanted the robot to act as their \textit{collaborator} in facilitating learning activities for young children. We therefore introduce this notion of \textit{parent-robot collaboration} to capture the collaborative dynamics between the parent and the robot. Furthermore, we identified three dimensions that determine various possibilities of parent-robot collaboration scenarios and will later refer to them jointly as \textbf{CAM} factors, which represents parent/robot relative capability (C), parent availability (A), and parent motivation (M). We believe that CAM helps inform a desired parent-robot collaboration dynamic, guiding decision regarding the robot's responsibilities based on practical situations. Additionally, educational objectives and concerns are the two main parental factors that define and constrain parent-robot collaboration respectively. These two factors are connected to some of the strategies, especially Restrictive Mediation, described in the parental mediation theory from prior literature (Section~\ref{sec-2.3}). Our findings regarding parents' concerns about invasion of privacy and decreased human-human interaction for children also interestingly align with concerns teachers raised for students using a social robot in a classroom setting {\cite{van2020teachers}}. Finally, the concepts of interaction dynamics and interaction initiatives articulate the interaction with the robot in terms of the number of participants and the timing of interaction, implying that the interaction design should support flexible multiparty scenarios and mixed-initiatives \cite{horvitz1999principles}. In the following sections, we unpack the connections between these themes to elaborate the notion of parent-robot collaboration and explain how they contribute to the contextual understanding of interaction design for a learning companion robot through the lens of parental mediation theory \cite{valkenburg1999developing, beneteau2020parenting} and mixed-initiative principles \cite{horvitz1999principles}. 

\subsection{Research Implications}\label{sec-5.1}

We see an opportunity to extend our understanding of parent-robot collaboration and the factors that might affect the emergence and outcomes of this collaboration by connecting findings across the three themes emerged in our analysis, which we discuss below.

\subsubsection{\textbf{Robot's responsibilities in parent-robot collaboration:} Parent-guided and CAM factors}

To determine what the robot should or should not do when collaborating with a parent to support a child in a learning context, the connection between the \textit{parent-guided factors} (Section~\ref{sec-4.2}) and the \textit{CAM factors} (Section~\ref{sec-4.1}) plays an important role. From an overarching perspective, parents' \textit{educational objectives} and \textit{concerns} determine what the robot should support and restrict. This aligns with the idea of Restrictive Mediation from Parent Mediation Theory, referring to how parents make rules to guide or prohibit the use of technology for children \cite{valkenburg1999developing}.  Meanwhile, from a practical standpoint, \textit{CAM factors} further determine specific tasks for the robot based on real-life scenarios. For instance, in our study, P1 wanted the robot to help her child understand processes of problem solving (\textit{educational objective}). However, P6 was hesitated to engage the robot due to doubts about the sources of the robot's content (\textit{concerns}), affecting her willingness to trust the robot's responses. The example also implies that the robot's overarching responsibility should include not only supporting the child's learning but also explaining sources of its content to mitigate parents' concerns. 

While parents could leverage trustworthy content sources to support their child's understanding of problem solving, the \textit{CAM factors} may practically limit parents from carrying out the tasks by themselves, creating opportunities for the robot to support and thereby defining the practical responsibilities the robot should take on. For instance, some parents may not have sufficient pedagogical knowledge (\textit{C}) to effectively support a child's problem-solving process, but they may have the time (\textit{A}) and strong motivation to do so (\textit{M}). In such cases, the robot's responsibility would be to compensate for the parents' lack of pedagogical knowledge as needed, rather than completely taking over the task. Conversely, some other parents may have sufficient pedagogical knowledge (\textit{C}) and motivation (\textit{M}) but may be too busy (\textit{A}) to support their children. In such cases, if the robot has better capability than the parent, it could take over the job during parental absence. However, if the robot does not have the same or better level of capability as the parent, it is worth considering how the parent could transfer their knowledge to the robot so that the robot could later support the child independently. These examples demonstrate that we can use \textit{parent-guided factors} to establish the overarching goals the robot needs to help achieve and identify the concerns the robot needs to help mitigate. Additionally, we can utilize \textit{CAM factors} to further consider parents' participation and ultimately determine the tasks the robot should and should not undertake in real-life scenarios.

\subsubsection{\textbf{Robot's level of involvement in parent-robot collaboration:} Robot-guided and CAM factors}

To collaborate with parents to support children in learning contexts, the robot may (1) take over the responsibility to support the child in a \textit{child-robot interaction} scenario, (2) at times adjust its level of involvement dynamically in a \textit{parent-child-robot interaction} scenario, or (3) exclude itself from the interaction to support a \textit{parent-child interaction} scenario. The interaction dynamic among the three agents (i.e., parent, child, robot) shifts across different patterns from time to time based on parent-child needs. We note that the idea of parent-robot collaboration should extend beyond interactions involving a parent, a child, and a robot, taking a holistic view and including scenarios where the robot helps carry out the parenting practices parents would have done on their own but were unable to carry out practically. The connection between the \textit{CAM factors} and the \textit{interaction dynamics} (see Section~\ref{sec-4.3}) provides insights into various interaction dynamics the robot might need to handle. 

In the previous section, we described how \textit{CAM factors} influence the delegation of responsibility between a parent and the robot. Here, we want to emphasize how the outcome of this delegation naturally shapes diverse interaction dynamics the robot needs to manage. For instance, P7 suggested that when she was occupied with other kids (A), she preferred the robot to engage solely with her child, forming a \textit{robot with a single child} dynamic; other times she preferred involvement in a \textit{robot, parent, and child} dynamic. In addition, the three parties involved in a parent-child-robot triadic interaction have varying levels of involvement based on CAM factors. For example, P3 wanted the robot to provide prompts for her to lead the conversation due to her strong teaching ability. In this case, the parent preferred more involvement than the robot. In contrast, P7 appreciated the robot taking the lead, with her providing supplementary support given her time constraints. In this case, the parent preferred the robot to be more involved than herself. Finally, P13 explained that she thought it was valuable to sometimes have one-on-one time with her child in a \textit{parent-child} bonding opportunity without the robot involved. These examples show how we can use \textit{CAM factors} not only to explain what the robot should do but also the interaction dynamics the robot should handle from time to time.

\subsubsection{\textbf{Robot's initiative in parent-robot collaboration:} Parent-guided and robot-guided factors}

In the process of parent-robot collaboration, the robot may take automated initiatives for the interaction or be directly controlled by the participants (e.g., the parent or the child). The parents' preferences for the forms of initiatives could be described by the connection between \textit{parent-guided factors} (Section~\ref{sec-4.2}) and \textit{interaction initiatives} (Section~\ref{sec-4.3}), as well as the type of activities in which participants are involved. For example, P7 explained that she preferred certain informal learning activities that required creativity and idea generation to grant the robot a higher degree of autonomy in taking initiatives, but only during a specific period of time. P3, on the other hand, preferred direct control over the robot as she had her own vision of how to guide the child in learning, using the robot as a supportive tool. P4 took a unique perspective of authoring the robot's initiatives and allowing the robot to autonomously execute her decisions. 

We reflect on this concept of interaction initiatives with the notion of mixed-initiatives proposed by \citet{horvitz1999principles}, which discussed mixed-initiative principles that aim to combine the power of direct user operation and system automation services to improve the user experience and foster productivity during collaborative work. They argued that a mixed-initiative system should gather information and make inferences about users' intentions and needs to properly handle uncertainty during the interaction. In addition, they demonstrated the application of mixed-initiative principles to task-oriented computer systems, which mainly support computer-based tasks to increase productivity in general workspace \cite{horvitz1999principles}. Our findings contribute to an extension of this notion to relatively complicated home environments, physically-present embodied agents, as well as the nature of a multi-party interaction dynamic, which involves the different needs of each individual as well as the parenting policies that govern and constrain the child's use of the robot. 

\subsection{Design Implications}\label{sec-5.2}

Here, we present the design considerations for the notion of \textit{parent-robot collaboration}. We note that this idea extends beyond interactions actually involving a parent, a child, and a robot. Instead, we adopt a holistic perspective and include scenarios where the robot helps implement the parenting practices that parents would have done by themselves but were unable to carry out practically. In the upcoming sections, we articulate the meaning of each design implication and outline potential design considerations for practical implementation. Based on the discovery of three major themes and their connections, we have identified the following design implications:

\subsubsection{Design Implication \#1: } \textbf{To enable parent-robot collaboration, a learning companion robot should align tasks with parental goals, address concerns, and supplement parental limitations based on capability, availability, and motivation.} This implication addresses what tasks the robot should handle in collaboration with parents. Most of these decisions rely on parents' intentions, which, in turn, depend on user information. Following the concept of mixed-initiatives, the robot should either seek direct input from parents before or during the interaction or make intelligent inferences by monitoring user activities. Moreover, the robot should effectively communicate its capabilities during collaboration. For example, it can inform the parent if it cannot detect the child's emotional state or provide specific personalized support. This communication helps establish accurate user expectations regarding the robot's capabilities and limitations.  

\subsubsection{Design Implication \#2: } \textbf{To enable parent-robot collaboration, a learning companion robot should predict and adapt interaction dynamics based on user preferences and ongoing user input.} This implication emphasizes the varying levels of engagement among participants in the interaction and highlights the need for the robot system to possess technical capabilities to accommodate diverse interaction dynamics based on individual user requirements. Additionally, parents may alter their participation by either rejoining the interaction later or temporarily leaving to address urgent matters. Therefore, future design should consider the robot's adaptability to changing dynamics and the seamless integration of ongoing user feedback.

\subsubsection{Design Implication \#3: } \textbf{To enable parent-robot collaboration, a learning companion robot should consider parenting policies and contextual needs when determining its interaction initiatives.} This implication addresses the timing and circumstances under which the robot should take the initiative in an interaction. The degree of autonomy granted to the robot should be determined not only by established parenting policies but also by ongoing parental input. This approach helps refine the robot's understanding of user preferences and contextual needs over time. Additionally, the robot should be equipped with the technical capabilities necessary to support mixed-initiatives. For instance, the robot may need to be capable of initiating a conversation without interrupting an ongoing interaction. Conversely, there may be times when the robot needs to switch to a more passive mode, allowing users to have full control over its initiatives. In future design, it's important to consider methods for users to communicate their preferences for interaction initiatives with the robot. Additionally, models should be developed that enable the robot to adapt to different modes of initiatives smoothly.


\subsubsection{Integrating Design Implications in Practice}

To implement the design implications, we propose the following preliminary interaction design workflow, which integrates our implications and includes concrete examples of potential features. Parents should have the ability to continuously adjust and add their preferences to the robot as they progress through each of the following steps. It's important to note that these are early-stage concepts based on qualitative evidence. Future work is needed to develop prototypes and evaluate these design concepts and technical requirements.

\begin{enumerate}
    \item \textbf{Set up background profile for both the parent and the child: } Participants create profiles to share demographic information and personal preferences, enabling the robot to customize interactions, such as recommending age-appropriate content. During this step, clear explanations about data storage and usage should be provided, while also respecting user privacy preferences and offering the option to prevent sharing non-essential information to address privacy concerns.
    \item \textbf{Define learning objectives and concerns: } The robot can prompt parents to define their learning objectives and concerns, collaboratively generating feasible and acceptable tasks to support the child's progress. The system may auto-suggest ideas for task definition, validate user choices, offer guidance and recommendations to accomplish each task, and highlight potential risks and benefits.  
    \item \textbf{Define task delegation between parent/robot: } After tasks are defined, the robot can propose task distribution between the parent and itself by collecting data on CAM factors from users: parental capability (e.g., confidence in task execution), availability (e.g., when they can participate), and motivation (e.g., time commitment). Parents should have the ability to modify the task assignment, with the robot transparently acknowledging its limitations to establish realistic user expectations.
    \item \textbf{Suggest and confirm preference for interaction dynamics: } As CAM choices are determined, the robot can display a visual report illustrating the potential interaction dynamics between the parent and the robot (e.g., a pie chart representing their respective contributions). This allows the parent to assess their desired level of involvement. 
    \item \textbf{Suggest and confirm preference for interaction initiatives: } The robot can propose various approaches to initiate the interaction during collaboration, seeking confirmation from the parents to ensure alignment with their expectations.
    \item \textbf{Save desired interaction paradigms: }Finally, we suggest that parents have the option to save frequently used interaction paradigms, creating distinct \textit{personas} for the robot that can support collaboration across various objectives. 
\end{enumerate}

\subsection{Limitations \& Future Work}\label{sec-5.3}

A key limitation of our work is that our participant sample offers limited representation in the following socio-demographic dimensions: (1) \textit{location:} all participants lived in the Midwestern United States; (2) \textit{gender:} all parents were females; (3) \textit{income:} more than half of the families have middle to high level of household income; and (4) \textit{education:} more than half of the mothers had a Master's degree or above. First, the parents self-selected which parent would participate, rather than the study team, and thus our entire sample was composed of mothers. This is consistent with prior literature in education \cite{schoppe2013comparisons, mcbride1993comparison}. Despite our efforts to recruit a representative sample, it was challenging to balance multiple demographic factors (e.g., gender, race, income, maternal education, etc.) all at the same time. Plus, fathers at times signed up for the study but mothers ended up being the ones who participated. Similarly, due to common challenges of recruiting lower-income families {\cite{nicholson2011recruitment}}, we have little low-income families in our sample. It is well-established that children in families with higher income or education levels tend to have better academic abilities {\cite{sirin2005socioeconomic}}, which may influence the kinds of support those families prefer the robot to offer. There may be specific needs and support that fathers or lower-income families prefer but were not well captured in our results because of lacking enough representatives in our sample. We will expand  our efforts to recruit a more representative population in our future work.


We conducted brief home visits to gain insight into parents' perspectives on using a learning companion robot. However, this short-term exploration may restrict our findings to the short-term effects and immediate parental experiences with the robot. It may not encompass the evolution of interactions over time, necessary updates, and design considerations for adapting to changes within an evolving ecosystem. For example, parents have voiced concerns regarding content appropriateness and the need for parental control. These sentiments may evolve over an extended period as parents develop stronger bonds with the robot and the robot demonstrates its reliability. Future research should employ long-term deployment approaches to examine how parents engage with the robot over an extended duration. 

Finally, our work has presented qualitative findings concerning user perceptions, preferences, and design implications. To practically implement our design recommendations and better assess their feasibility, it is essential to construct viable prototypes and analyze the statistical relationships between the identified factors. Future research should prioritize the development of prototypes based on the design implications, evaluate the practicality of these design concepts, and quantitatively investigate the relationships between the factors.

\section{Conclusion}

In this paper, we conducted an in-home study with 10 parent-child pairs, each with a child aged 3--5 years old, to explore their interaction with a learning companion robot in an informal learning setting. We developed prototype capabilities for a learning companion robot to deliver educational prompts and responses to the parent-child pairs during reading sessions. Our study data revealed that parents wanted to \textit{collaborate} with the robot to support their child's learning, introducing the concept of \textit{parent-robot collaboration}. We further unpacked factors influencing shared responsibility between the parent and the robot, guiding or limiting the use of the robot, and shaping interaction dynamics and initiatives enabled by the robot. Our findings offer an empirical understanding of the needs and challenges of parent-child interaction in informal learning activities, along with design prospects for incorporating a companion robot into such scenarios. We contribute valuable guidance on how robots can be designed to foster parent-robot collaboration, encompassing supplementing practical limitations of parents, augmenting parenting policies, and adapting to flexibal interaction paradigms.

\begin{acks}
This work was supported in part by the National Science Foundation awards 2202803 and 2312354. We thank Nathan White for his help with the technical development of our robot system and Jungseok Roh for his help with data collection during the field studies.
\end{acks}

\balance
\bibliographystyle{ACM-Reference-Format}
\bibliography{Bibliography}


\end{document}